\documentclass[conference]{IEEEtran}

\IEEEoverridecommandlockouts

\usepackage{cite}
\usepackage[bookmarks=false]{hyperref}
\usepackage{amsmath,amssymb,amsfonts}
\usepackage{algorithmic}
\usepackage{graphicx}
\usepackage{float}
\usepackage{subcaption}
\usepackage{textcomp}
\usepackage{xcolor}
\usepackage{url}
\usepackage{listing}
\usepackage[frozencache=true,cachedir=_minted-main]{minted}
\usepackage{csvsimple}
\usepackage[normalem]{ulem}

\graphicspath{{../Figures/}}
\DeclareGraphicsExtensions{.pdf,.jpeg,.png}

\def\BibTeX{{\rm B\kern-.05em{\sc i\kern-.025em b}\kern-.08em  T\kern-.1667em\lower.7ex\hbox{E}\kern-.125emX}}

\begin{document}

\title{The BioExcel methodology for developing dynamic, scalable, reliable and portable computational biomolecular workflows}

\author{
\IEEEauthorblockN{Jorge Ejarque\IEEEauthorrefmark{1},
Pau Andrio\IEEEauthorrefmark{1},
                  Adam Hospital\IEEEauthorrefmark{2},                  Javier Conejero\IEEEauthorrefmark{1},\\
                  Daniele Lezzi\IEEEauthorrefmark{1},
                  Josep LL. Gelpi\IEEEauthorrefmark{1}\IEEEauthorrefmark{3},
                  Rosa M. Badia\IEEEauthorrefmark{1}}
\IEEEauthorblockA{\IEEEauthorrefmark{1}Barcelona Supercomputing Center (BSC-CNS), Barcelona, Spain }
\IEEEauthorblockA{\IEEEauthorrefmark{2}Institute for Research in Biomedicine (IRB), Barcelona, Spain}
\IEEEauthorblockA{\IEEEauthorrefmark{3}Dept. Biochemistry and Molecular Biomedicine, University of Barcelona, Spain}

jorge.ejarque@bsc.es, pau.andrio@bsc.es,  adam.hospital@irbbarcelona.org, francisco.conejero@bsc.es,
\\
daniele.lezzi@bsc.es,  josep.gelpi@bsc.es,
rosa.m.badia@bsc.es
}

\maketitle
\thispagestyle{plain}  
\pagestyle{plain}      


\begin{abstract}
Developing complex biomolecular workflows is not always straightforward. It requires tedious developments to enable the interoperability between the different biomolecular simulation and analysis tools. Moreover, the need to execute the pipelines on distributed systems increases the complexity of these developments. To address these issues, we propose a methodology to simplify the implementation of these workflows on HPC infrastructures. It combines a library, the BioExcel Building Blocks (BioBBs), that allows scientists to implement biomolecular pipelines as Python scripts, and the PyCOMPSs programming framework which allows to easily convert Python scripts into task-based parallel workflows executed in distributed computing systems such as HPC clusters, clouds, containerized platforms, etc. Using this methodology, we have implemented a set of computational molecular workflows and we have performed several experiments to validate its portability, scalability, reliability and malleability.
\end{abstract}


\section{Introduction}
\label{sec:introduction}
Computational workflows are one of the most used tools to assemble and run simulations of different scientific fields as climate predictions, bioinformatics, engineering, etc. Researchers can compose their applications, usually made of pieces of code available in libraries and binaries, using a textual or graphical representation of the dependencies between those parts, and let the runtime of the workflow management system to orchestrate the execution on a given computational platform. In particular, HPC systems are becoming more attractive for the execution of workflows that traditionaly have been executed on distributed systems as grids or clouds, because they can have tasks that require a certain degree of massive parallelism (i.e., OpenMP/MPI tasks, GPUs). The trend is to have complex HPC systems built on hybrid architectures that combine traditional processors with accelerators and other devices. On top of the computing complexity, the packaging of workflows is an additional issue, with containers becoming a popular way to distribute and deploy applications.

To address the issues above, it is a must to have a system that can offer a simple interface for the composition of applications components which are able to transparently manage their execution, adapting them to the different capabilities of heterogeneous computing and to the dynamicity of the computational load.Moreover, the computation requirements of these workflows impose to scale their execution to a large amount of resources and to provide reliability mechanisms. 

This paper presents a methodology for defining and orchestrating biomolecular simulations on HPC infrastructures that satisfies the above mentioned requirements. It combines a software library developed by the BioExcel Centre of Excellence, BioExcel Building Blocks (BioBBs)~\cite{web:biobb}, with a workflow programming framework, PyCOMPSs~\cite{compss_softwareX}~\cite{pycompss}. On the one hand, the BioBB library allows scientist to implement pipelines as Python scripts that automatize the various steps of Molecular Dynamics (MD) simulations that are performed manually in many cases. On the other hand, PyCOMPSs converts Python scripts into task-based parallel workflows and orchestrates the execution of the computational tasks in resources of distributed computing systems such as HPC clusters, clouds or containerized platforms~\cite{pycompss2}. Using this methodology, we have implemented a set of computational molecular workflows and performed several experiments to validate its portability, scalability, reliability and elasticity.

The paper is structured as follows: Section~\ref{sec:biobbs} describes the proposed methodology; Section~\ref{sec:workflows} presents the workflows implemented adopting this methodology; Section~\ref{sec:validation} reports the validation of the features provided by the presented methodology; Section~\ref{sec:related_work} presents the state of the art and related work on topics involved in the proposed research. Finally, Section~\ref{sec:conclusions} draws the conclusion and proposes guidelines for future research in this topic.

\section{Related Work}
\label{sec:related_work}
The use of computational workflows has become ubiquitous for data analytics in the field of bioinformatics since the last decade. In the literature, more than 200 workflow systems~\cite{web:wflist} can be found, targeting specific scientific domains, different execution models and usability approaches. Workflow systems can be classified according to the model used to define the tasks and the data dependencies and to the characteristics of the engine that executes the workflow on the computing platform. With relation to the tasks definition features, some frameworks allow to explicitly define the workflow through a recipe file or a graphical interface while others permit the users to program their applications and let the runtime build a dependency graph from the user code. Another relevant characteristic for the classification of these frameworks is the level of integration with the different computing platforms as distributed environments (such as grids, clouds, and clusters), and HPC systems with multi-core architectures and accelerators (such as GPGPUs).

Amongst all these tools, in this paper we focus on the features that are more convenient for the orchestration of molecular dynamics simulations, taking into account interoperability across a variety of software and hardware environments, scalability, and reproducibility. In particular, we consider HPC-focused workflow managers that can compose and run workflows with advanced features as elasticity, adaptability, and fault tolerance.

Taverna~\cite{taverna, web:taverna}, Kepler~\cite{kepler, web:kepler}, Galaxy~\cite{galaxy, web:galaxy} are well known graphical environments for the composition of workflows that can be stored and shared with other users of the community. These graphical environments facilitate the design of simple workflows, but the implementation of complex dynamic algorithms is difficult.

Other frameworks provide more generic interfaces to express the components of the workflow. Toil~\cite{toil} is a Python workflow engine focused on the execution of pipelines. Pipelines are defined as jobs that can contain children jobs and follow-on jobs to explicitly force the synchronization of the execution. Nextflow~\cite{nextflow, web:nextflow} provides a fluent DSL to implement and deploy scientific workflows and allows the adaptation of pipelines written in the most common scripting languages.
 


More bioinformatics specific environments have been recently developed.
Crossbow~\cite{web:xbow} is a Python-based toolkit for workflow construction and execution, aimed mainly at Crossbow clusters but more generally at distributed computing environments. It provides an easy entry to cloud-based computing for biomolecular simulation scientists. Crossbow shares many of its design aspects with Parsl~\cite{parsl}. It provides tools to wrap Python functions and external applications (e.g., legacy MD simulation codes), in such a way that they can be combined into workflows using a task-based paradigm. Crossbow uses Dask~\cite{dask} Distributed as the task scheduling and execution layer.

RADICAL-Cybertools~\cite{Balasubramanian2019RADICALCybertoolsMB} enables the execution of ensemble-based applications on a variety of high performance computing infrastructures. An increasing number of scientific domains are adopting and benefiting from ensemble-based applications. Most notably, MD simulations are nowadays executed as many parallel jobs of ns-length simulations rather than a single, long, and very large MPI job.
AdaptiveMD~\cite{web:adaptivemd} is a Python package designed to create HPC-scale workflows (parallel tasks) for adaptive sampling of biomolecular MD simulations. AdaptiveMD is designed as a distributed application that can be launched from a laptop or directly on an HPC resource and asynchronously automates the workflow creation and execution. Multiple adaptive sampling algorithms are fully automated with minimal user input, while advanced users can easily make modifications to workflow parameters and logic through the Python API. Runtime adaptations include the use of interim data as task properties such as analysis types or parameters, and workload properties such as task count or convergence criteria. To provide robust workflow management, AdaptiveMD is also integrated with the RADICAL Cybertools stack, which significantly enhances the runtime error detection and correction functionality, but has a much higher installation and configuration overhead.

The solution described in this paper advances the mentioned
approaches in the move to developing robust and scalable scientific workflow without the requirement of deep programming knowledge on the users. The adoption of the combination of BioExcel BioBB and PyCOMPSs provides powerful features which simplify the development and executions of complex bio molecular workflows combining several types of heterogeneous tasks running in parallel on thousands of computing cores. Graphical workflow systems like Galaxy and KNIME have generally limited support for using HPC and HTC compute infrastructure in combination with high-performance codes like GROMACS, while our solution provides a solid solution for the execution of applications on a lot of computing backends without the need of adapting the code to a specific one.


\section{Methodology}
\label{sec:biobbs}

\begin{figure*}[tb]
\centering
    \begin{subfigure}[b]{0.50\textwidth}
        \centering \includegraphics[width=\textwidth]{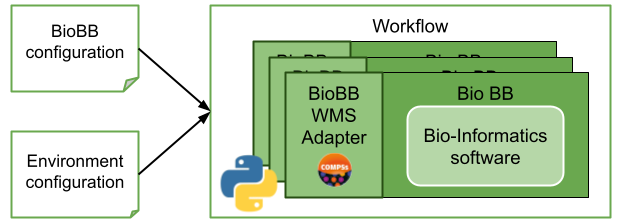}
        \caption{Development}
    \end{subfigure}
    \hfill
    \begin{subfigure}[b]{0.48\textwidth}
        \centering     \includegraphics[width=\textwidth]{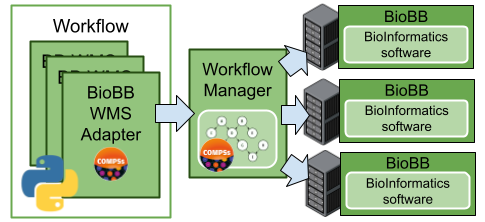}
        \caption{Execution}
    \end{subfigure}
    \caption{Methodology overview}
    \label{fig:overview}
\end{figure*}

Joining different biomolecular tools in a complex pipeline is not always straightforward. It requires tedious developments to enable the interoperability between the different biomolecular simulation and analysis tools. Moreover, the need to execute the pipelines on distributed systems increases the complexity of these developments. Figure~\ref{fig:overview} provides an overview of the methodology proposed by the BioExcel Center of Excellence to simplify the development of dynamic, scalable, reliable and portable computational biomolecular workflows for distributed computing infrastructures. With this methodology, workflows are developed as simple Python scripts using reusable and extensible modules. The inputs of these workflows are two configuration scripts: one to configure the different workflow modules to setup the biomolecular system to be evaluated (mutations, number of simulation steps, etc.) and another one to indicate the properties related to the specific computing infrastructure for a specific execution (number of cores per node, gpus, etc.). The modules used inside the workflow are composed of two layers: the first layer is provided by the BioExcel Building Blocks (BioBBs), a software library designed to tackle the interoperability problem thanks to a simple wrapping approach. BioBBs are a collection of small wrappers written in Python where each building block encapsulates software components and provides a well-defined interface for input, output, configuration, and provenance. A standardised syntax is used in all the building blocks, with each of the wrappers internally performing the necessary format conversions for input and output, and launching the tool which runs unaltered. With this design, a large set of biomolecular tools such as GROMACS~\cite{10.1093/bioinformatics/btt055} (MD), HADDOCK~\cite{haddock} (Docking) or PMX~\cite{pmx2015} (Free energy calculations) can be executed using an homogeneous syntax, also providing a uniform and stable interface with enough information to plug the components into interoperable workflows as simple Python scripts. To transparently integrate the BioBBs with Workflows Management System, we propose a BioBB WMS Adapter layer, which is implemented as a set of decorators which interacts with the management system that takes care of the execution of the workflows. This adaptor layer transforms the local calls to the BioBBs into remote asynchronous calls. In this work, we have used PyCOMPSs to implement the adaptation layer. PyCOMPSs provides a programming model and a runtime system which allows developers to easily convert a sequential Python script to parallel workflows for distributed computing environments hiding the complexity of the parallelization and of the execution management. Next paragraphs provide more details about how this methodology provides different functionalities.

\subsection{Programmability, reusability and portability}
\begin{figure*}[ht!]
    \centering
    \includegraphics[width=1\linewidth]{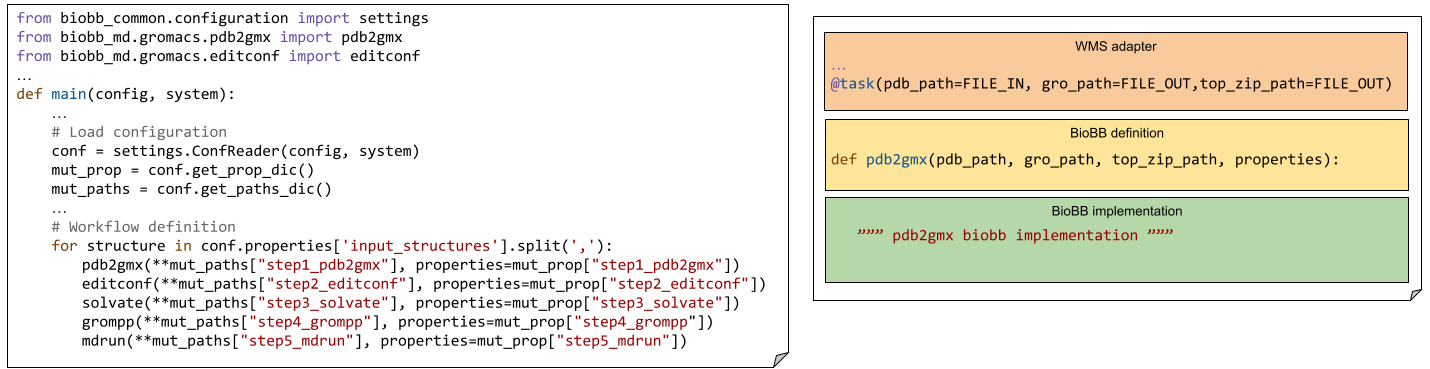}
    \caption{Workflow implementation example. The left-hand side of the picture shows a code snippet of a workflow invoking different BioBB as standard python functions. The right-hand side of the picture shows how a BioBB and its WMS adapter is implemented.}
    \label{fig:workflow}
\end{figure*}
Figure~\ref{fig:workflow} shows a code snippet to explain how workflows are developed with this methodology. On the left side of the figure, we can see a molecular dynamics setup workflow developed as a simple Python script where the executions of the computational biology tools are modelled as invocations to standard Python methods. These methods are provided by the BioBB libraries so developers do not need to implement them. They only need to import the modules of the tools they want to use in their workflows. In the right side of the picture, we can see a snippet of this BioBB module. The Python method definition is used as a common interface where the first parameters indicate required data inputs and outputs, and the properties parameter is a Python dictionary indicating the configuration of the execution of the tool. The body of the Python method contains the implementation while the decorators on  top of the function are the PyCOMPSs annotations that enable the parallelization and execution of the BioBBs modules in the different computing nodes. All the BioBB methods are annotated with the task decorator which is also used to indicate the direction of the BioBB parameters. Every time a BioBB is invoked in a workflow, a task is created by the PyCOMPSs runtime that analyses the dependencies between the different BioBB according to the indicated direction. If the task is free of dependencies the BioBB execution will be scheduled and asynchronously executed in a computing resource. In this way, PyCOMPSs hides the complexity of the parallelization and the execution distribution of the workflow. Moreover, note that the workflow code is infrastructure agnostic, because the developer does not need to specify details of the infrastructure in the code that can be executed in different infrastructures without modifications. %

\subsection{Multi-level parallelism}
Some computational biology tools are internally parallelized either to use different CPU cores in a single node (such as threads or OpenMP), across multiple nodes (such as MPI) or as a combination of both (MPI+openMP). It must be indicated in the workflow manager system in order to allocate the necessary resources for the execution. In our approach, this information is provided by adding decorators in the adapter part of the BioBB (\textit{@multinode} and \textit{@constraint}). Figure~\ref{lst:task_level} shows a code example  in the case of a GROMACS mdrun simulation. The \textit{multinode} decorator indicates that the execution is using several nodes and the \textit{constraint} decorator indicates the number of CPU or GPUs to use in each node of the execution. The values for these decorator are obtained from a set of environment variables defined in the environment configuration file of the workflow according to the computing platform capabilities. With this information, the PyCOMPSs runtime ensures the proper allocation of the required resources to perform the computation(via taskset, OMP\_NUM\_THREADS, CUDA\_DEVICE,...)

\begin{figure}[!htb]
  \centering
  \scriptsize
  \begin{minted}{python}

@multinode(computing_nodes = '$TASK_NUM_NODES')
@constraint(processors=[
  {'processorType':'CPU','computingUnits':'$TASK_NUM_CPUS'},
  {'processorType':'GPU', 'computingUnits':'$TASK_NUM_GPUS'}
  ])
@task(...)
def mdrun(...):
    ...

  \end{minted}
  \caption{Decorators to indicate different levels of task parallelism.}
  \label{lst:task_level}
\end{figure}

\subsection{Reliability}
Computational biomolecular workflows are long lasting analyses that perform large amounts of simulations. Some of these simulations could fail or take too much time to get to a solution. However, the kind of analysis performed in these types of workflows can reach a solution even if there are simulations that fail. To manage this kind of features, PyCOMPSs provides additional properties in the task decorator to indicate how to proceed if the computation of this task fails or takes longer than expected~\cite{task_failure}. These properties are included in the adaptor layer as shown in Figure~\ref{lst:reliability}. On the one hand, the \textit{time\_out} property indicates the maximum duration of the task. If a task is exceeding the indicated duration, its execution is cancelled and considered as a failure. On the other hand, the \textit{on\_failure} property indicates to the PyCOMPSs runtime what to do if a task of this type fails. In this case, the runtime will ignore the failure and the output data will be set to the specified default values (such as an empty file, None values,...)

\begin{figure}[!htb]
  \centering
  \scriptsize
  \begin{minted}{python}
@constraint(...)
@multinode(...)
@task(..., on_failure='IGNORE', time_out='$TASK_TIMEOUT')
def mdrun(...):
    ...
  \end{minted}
  \caption{Decorators to indicate reliability features.}
  \label{lst:reliability}
\end{figure}

Beside this functionality, we have also implemented a stop/restart mechanism to allow users to restart a workflow run without requiring to execute all the tasks again. This functionality is useful when executions exceed the wall clock limit of the queue system. To perform a safe stop, the runtime catches the signals sent by the queue system manager for cancelling the processes of the execution. When the signal is received, all the running and pending tasks are cancelled and the data generated by the already executed tasks is moved to its final location. When the execution is restarted and the workflow reaches the point to invoke a task, the adapter part of the BioBBs checks if the generated data is already available and that is not empty. If this data already exists, it skips the execution, otherwise, if it does not exist or it is empty it creates a task in the runtime to perform the missing computation.

\subsection{Malleability}

As stated in the introduction, the computing load of the workflows during the whole execution is not homogeneous; depending on the different phases, the workflow will use more or less computational resources. In task-based programming models like PyCOMPSs an application can be represented as a Direct-Acyclic-Graph (DAG) where nodes represent tasks and arrows data-dependencies between the tasks. This DAG also inherently stores information about the computational load required by the application at any point of the execution. The runtime can know which is the maximum achievable parallelism for a certain execution by analysing the generated graph and it can identify, by considering the available resources, whether the application has potential for further parallelism or it is under-utilising the current resources. For instance, when there are lots of dependency-free tasks pending for execution, the application could run faster by allocating more resources. However, when there are more resources than available ready tasks, we will under-utilise the system and some of the nodes will be idle or not running at the maximum capacity.

To overcome these situations, the runtime has an auto-scaling module which is able to scale up and down the computing resources used by the application according to its demands. During the application execution, the runtime generates profiling information about the previous tasks execution, including statistical information about the duration of each task. Combining this information with the task dependency graph, the PyCOMPSs runtime periodically estimates the remaining parallel workload ($PW$) as a sum of all the dependency-free task resource requirements ($R_{T_{ready_i}}$) multiplied by their mean execution time ($\overline{ET_{T_{ready_i}}}$).  

\begin{equation}
PW=\sum_{\forall T_{ready}}{R_{T_{ready_i}} \overline{ET_{T_{ready_i}}}} \label{eq:workload}
\end{equation}
 
In the same way, the runtime can estimate the current infrastructure capacity ($IC$) with the sum of all resource capabilities ($C_{Resource_i}$)  multiplied by the mean execution time to get a new resource from the infrastructure provider($\overline{RT}$).

\begin{equation}
IC=\sum_{\forall{Resource}}{C_{Resource_i} \overline{RT}} \label{eq:infrastructure_capacity}
\end{equation}
 
These metrics are useful to determine when to add or remove a resource. If the estimated parallel workload is higher than the infrastructure capacity during the time to create a resource, it indicates that the application has enough load to be sped-up with a new resource. On the contrary, if it is smaller, it means that the application is starting to under-use resources.

If the first situation is produced, the runtime contacts the infrastructure resource manager to request a new computational node and once the new resource is available, the runtime starts a worker process which spawns the execution of tasks in the new resource.

If the second situation is produced, the runtime has to decide which compute node is the best candidate to be removed. To do so, the runtime calculates the number of underused resources as the difference of the parallel load and the current infrastructure capacity. Then, it ranks the computing nodes by the capacity and the current running load (number of running tasks and estimated time to finish). Based on this rank, the runtime selects the node which contains the underused resources and which has the minimum running load. Once a node is selected, the runtime removes the node from the worker pool and once all the running tasks have finished, it contacts the infrastructure resource manager to release it.

\begin{figure}[!htb]
  \begin{center}
  \begin{subfigure}[a]{1\linewidth}
    \centering
    \includegraphics[width=1\linewidth]{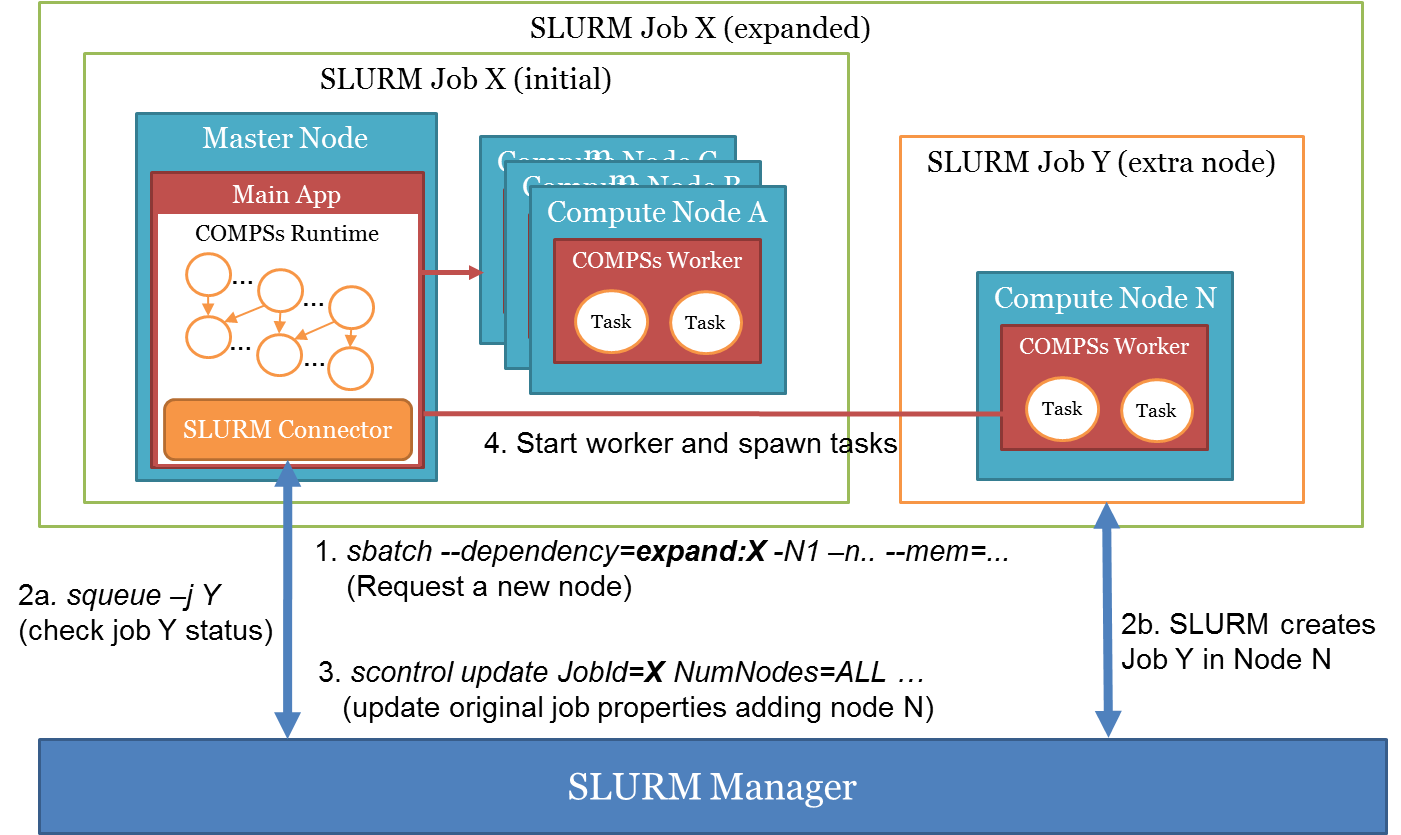}
    \caption{Resource scale-up}
    \label{fig:elasticty_sc_ups}
    \end{subfigure}
  \newline
  \begin{subfigure}[b]{1\linewidth}
    \centering
    \includegraphics[width=1\linewidth]{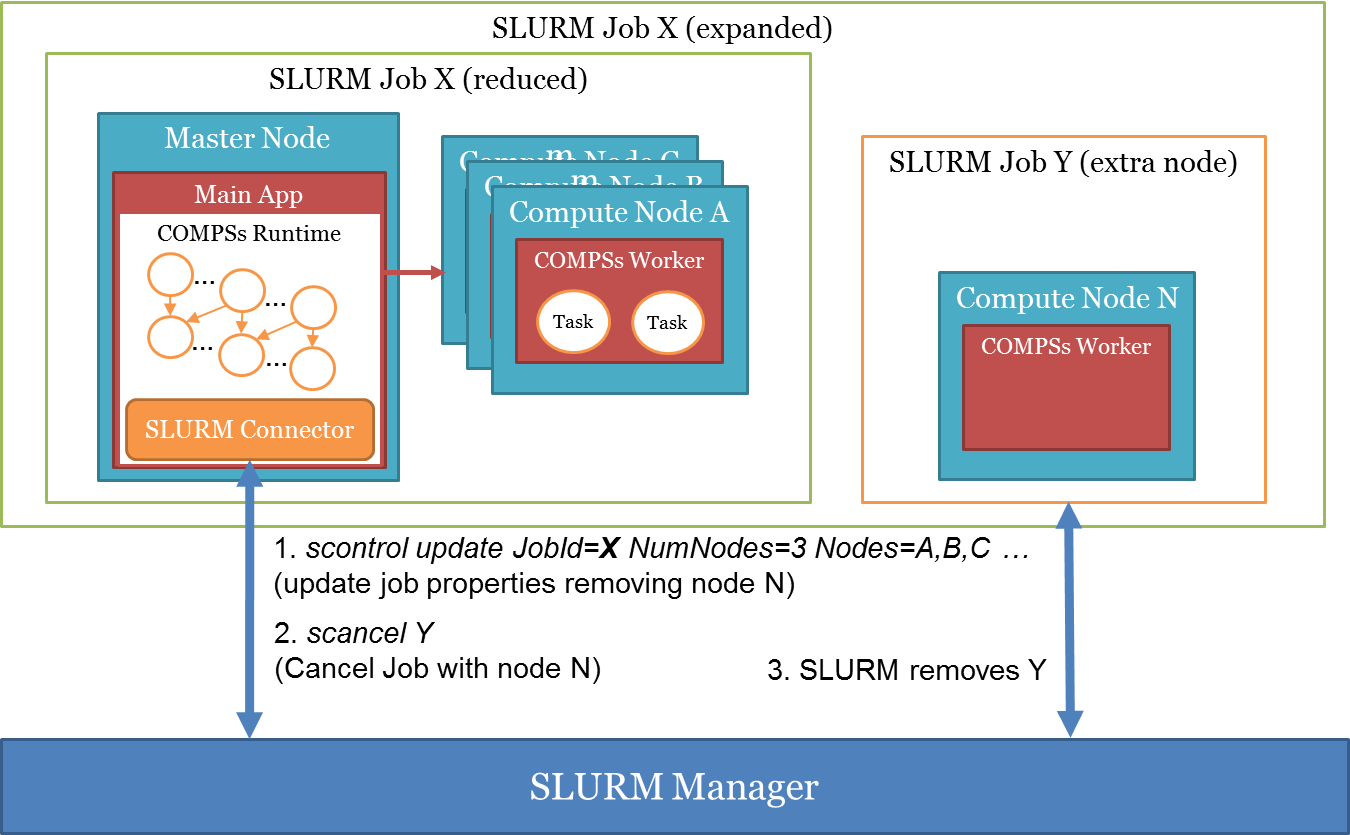}
    \caption{Resource scale-down}
    \label{fig:elasticty_sc_down}
    \end{subfigure}
 \end{center}
  \caption{Diagram about how the COMPSs runtime interacts with SLURM managed clusters to implement the resource elasticity in the workflow execution.}
  \label{fig:elasticity_slurm}
\end{figure}

In previous versions of PyCOMPSs~\cite{compss_servicess}, the auto-scaling features were applied to scale service executions in cloud environments, where the runtime contacts the resource provider API to create and destroy virtual machines. In this case, we have extended to auto-scale scientific workflows in clusters. Figure~\ref{fig:elasticity_slurm} illustrates how the runtime interacts with the SLURM resource manager to achieve the auto-scaling feature in clusters. To scale-up resources, the runtime requests an extra node by submitting a new job with the \textit{sbatch} command indicating with the dependency flag that this job expands the main application job. This job will inherit the same QoS from the expanded job and will execute a COMPSs worker process in the requested extra node. The runtime detects when the job is running by polling SLURM with the \textit{squeue} command. Once it is ready, the allocation of the new job is updated with the new resource with the \textit{scontrol update job} command. In the case of scaling-down, the runtime contacts the SLURM manager to cancel the job which contains the node to remove (with the \textit{scancel} command) and updates the main application job with the \textit{scontrol update job} command.


\section{Workflows}
\label{sec:workflows}
To validate the functionalities like infrastructure agnosticity, scalability, resiliency, and  malleability, we have implemented several workflows following the proposed methodology~\cite{web:biobb_workflows}. These workflows have been designed within the BioExcel CoE and applied to scientific use cases~\cite{use_cases} where we have evaluated different molecular systems. These workflows and systems are briefly introduced in this section.

\subsection{Mutations MD Setup Workflow}
\label{subsec:mdsetup}
\begin{figure}[!tb]
  \centering
  \includegraphics[width=.9\columnwidth]{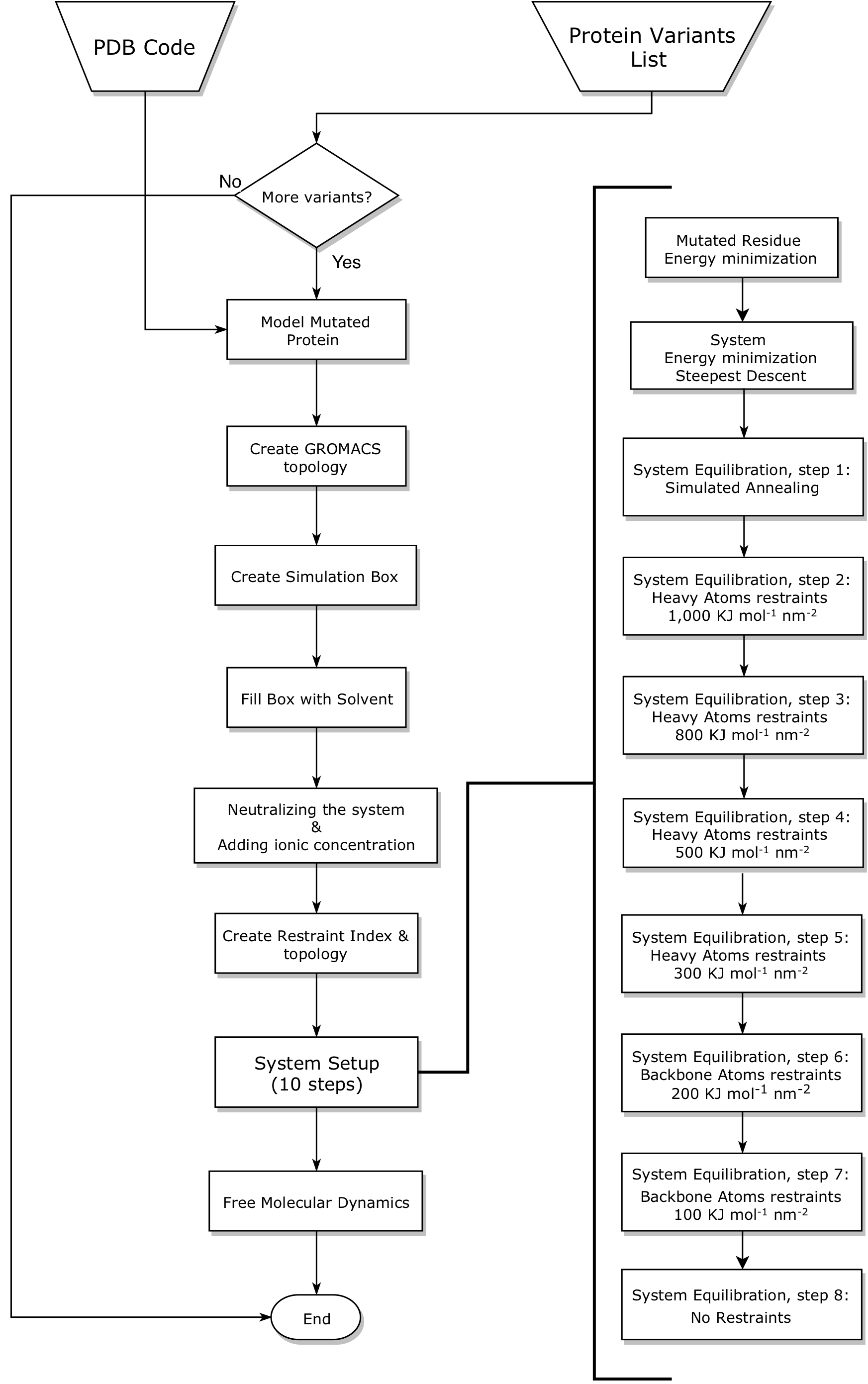}
  \caption{Mutation MD Setup flowchart}
  \label{fig:pymdflow}
\end{figure}

The Mutations MD setup workflow is an automated protocol to model residue mutations in 3D protein structures detected from genomics data, and prepare and run MD simulations for all the generated structures. The pipeline receives a PDB file (wild type protein 3D structure) and a set of mutations as input. Next, it prepares and runs MD simulations for each of the systems, thus obtaining static information (an ensemble of modelled structures for each of the protein variants), and dynamic data (trajectories for each of the protein variants). Both types of information can later be used in a comparative study. The workflow flowchart is represented in Figure~\ref{fig:pymdflow}. The structure of the workflow makes it a perfect case to study parallel work distribution, with a variable number of independent MD simulations to run, depending on the input number of mutations to model. Besides, the main tool used in the workflow (MD simulations) is implemented in various programming schemes, including GPU cards and openMP/MPI regimes, which makes the workflow also a perfect case to study the capacity to deal with different hardware architectures and parallelism levels. 

\subsection{Binding Affinity Workflow}
\label{subsec:pmxCV19}
\begin{figure}[!tb]
  \centering
  \includegraphics[width=.9\columnwidth]{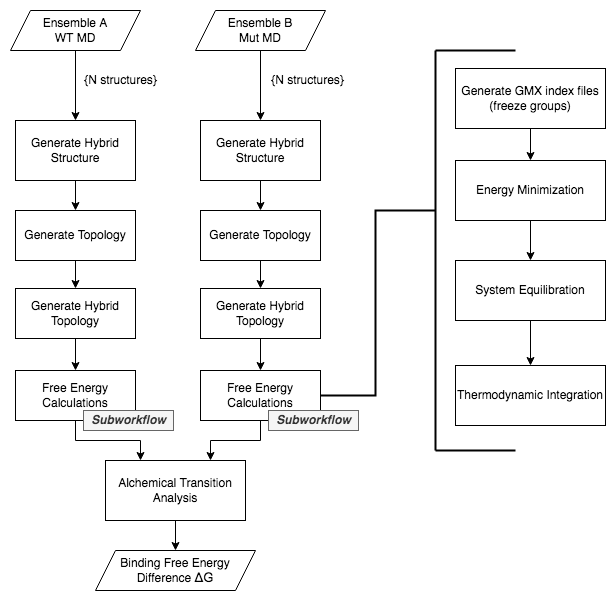}
  \caption{Binding Affinity flowchart}
  \label{fig:binding_affinity}
\end{figure}

The Binding Affinity workflow is an automated protocol to evaluate changes on binding affinity between a protein and its receptor upon residue mutations. The workflow uses the PMX tool~\cite{pmx2015} to generate alchemical residues, and GROMACS~\cite{10.1093/bioinformatics/btt055} to run a large series of short MD simulations required (thermodynamic integration, TI). The workflow flowchart is represented in Figure~\ref{fig:binding_affinity}. Briefly, the method starts with two trajectories (wild type, mutated) coming from equilibrium MD simulations. From these trajectories, two different ensembles of structures are extracted and used to prepare and run alchemical perturbations from the wild type protein to the mutated protein (forward) and from the mutated protein to the wild type protein (reverse). The final output of the workflow are two histograms built with the results of the forward and reverse simulations. The intersection between the two histograms defines the final $\Delta$G of binding.

The structure of the workflow makes it a perfect case to study PyCOMPSs fault tolerance, with a large number of complex independent workflow branches, with a certain probability of failure. In a typical run, 500 structures from each input ensemble are used to compute thermodynamic integrations, summing up to 1000 independent sub-workflows. The probability of failure for some of the workflow branches overall is high, but, unless the number of branches failing is high, it is not affecting the final value. Furthermore, this pipeline needs a considerable amount of computational resources and time, with the consequent risk of job cancelling, due to wall clock time or infrastructure downtime. Here the resilience properties to ensure the computation can be restarted and malleability properties to speed up the computation using more resources if available are very convenient.

\subsection{Molecular Systems}
Systems used in this study were chosen due to their interest in pharmaceutical research, and are briefly presented in this section.

The first system is the duodecimal peptide PMI, known to compete with p53 for binding to MDM2 or MDMX. p53 is critical for maintaining genetic stability and preventing cancer. MDM2 (E3 ubiquitin ligase) and its homologue MDMX act as negative regulators of p53. Designing inhibitors of MDM2 or MDMX is an attractive strategy for enhancing p53 activity and thus achieving the desired antitumoral therapeutic effect. The affinity of the peptide PMI is roughly two orders of magnitude higher than that of the same length p53 peptide. High-resolution crystal structures of both proteins with PMI are available (PDB identifiers: 3EQS and 3EQY, respectively). Kd dissociation constants of PMI and p53 for MDM2 and MDMX have been experimentally derived, and an additional Alanine scanning is also available for these proteins. All this experimental information can be used to test our methods to theoretically predict binding affinity.

The second system is the well-known SARS-Cov-2 Spike protein, and in particular, the molecules involved in the COVID-19 infection mechanism, the Spike Receptor Binding Domain (RBD) and the Human ACE2 (hACE2) receptor. High-resolution crystal structures of the protein complex are available (PDB identifier: 6VW1). Genetic changes in SARS-CoV-2 variants ($\alpha$, $\beta$, $\gamma$, $\delta$, $o$) are translated mainly into single amino acid mutations in the RBD domain of the viral Spike protein. The impact of these genetic changes can be determined using free energy calculations on the binding of RBD to the host receptor proteins. The system was chosen for its importance in the field, but also for its considerable complexity, both in size (~65000 atoms) and in the number of charged (ionizable) amino acids.

\section{Validation}
\label{sec:validation}
This section describes the experiments performed to validate the described capabilities and the analysis of the results. These experiments consist of the execution of the implemented workflows configurations in a set of supercomputers from the PRACE partnership and Spanish Supercomputing Network (RES). The description of these clusters are described in Table~\ref{tab:machines}

\begin{table}[htb]
\centering
\caption{Supercomputers description}
\label{tab:machines}
\begin{filecontents*}{Files/machines.csv}
Supercomputer, Computing Node Description
MareNostrum (MN), 48 core Intel Skylake CPU 
Minotauro (MT), 16 core Intel Haswell CPU + 4 Nvidia K80 GPU
Tirant, 16 core Intel Sandybridge CPU
Discoverer, 128 core AMD EPYC 7H12 CPU
\end{filecontents*}
\csvautotabular{Files/machines.csv}
\end{table}

\subsection{Portability}
To validate the portability features, we have executed the Mutations MD setup workflow with the MDM2-PMI Alanine Scanning system which executes 8 Molecular Dynamic pipelines using GROMACS. It has been executed in three supercomputers: MareNostrum (MN), Tirant, and MinoTauro (MT). Table~\ref{tab:portability} shows the results for the different executions. The first two columns show the selected supercomputer and the number of nodes used in each execution and the specified constraints for the MD task. Then, the third column shows the elapsed time to execute the whole workflow, and the last column shows the performance reported by GROMACS to compute the MD simulation. In this table we can observe that the workflow execution is adapted to the computing infrastructure reaching the expected GROMACS performance according to the available computing resources.
\begin{table}[htb]
\centering
\caption{Execution time in different supercomputers}
\label{tab:portability}
\begin{filecontents*}{Files/portability.csv}
Machine, MD Task Conf., Exec. Time, GROMACS perf.
MN (8 Nodes), 48 CPU cores, 97.5 min., 162.357 ns/day
MN (16 Nodes), 96 CPU cores, 64.0 min., 250.787 ns/day
Tirant (8 Nodes), 16 CPU cores, 372.8 min., 39.546 ns/day
MT (2 Nodes), 4 CPU + 1 GPU, 197.5 min.,  77.650 ns/day

\end{filecontents*}
\csvautotabular{Files/portability.csv}
\end{table}

\subsection{Scalability}
\begin{figure}[htb]
    \centering
    \includegraphics[width=1\linewidth]{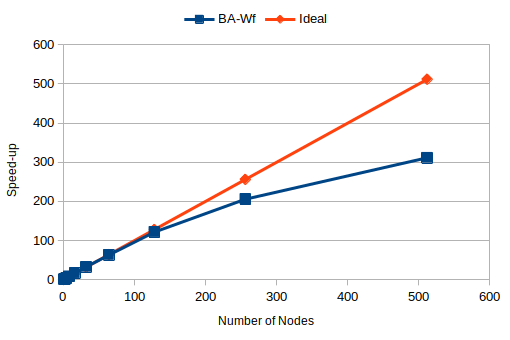}
    \caption{Strong scaling analysis for Binding Affinity Workflow. (The baseline is the execution time using a single node)}
    \label{fig:pmx_strong_scaling}
\end{figure}

To validate the scalability of the methodology, we have performed a strong and weak scaling analysis for the Binding Affinity workflow using the SARS-Cov-2 Spike protein system in the Dicoverer supercomputer.  Figure~\ref{fig:pmx_strong_scaling} shows the results of the strong scaling experiment. In this experiment, the workflow has been configured to evaluate 512 structures for each forward and reverse ensembles. The same workflow configuration has been executed with a different number of resources using a Discoverer node (128 cores) for each simulation. For each workflow execution, we have measured the execution time of the workflow computation discarding the runtime initilization and finalization, and we have computed the speed-up. The results show a good scalability up to 512 nodes (65,536 cores). It is close to the ideal until 256 nodes and reaches a speed-up of 311 for 512 nodes.  

\begin{figure}[htb]
    \centering
    \includegraphics[width=1\linewidth]{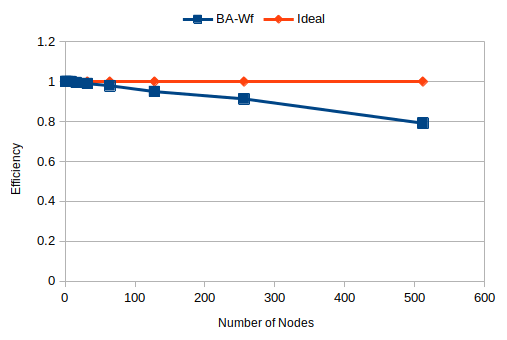}
    \caption{Weak Scaling analysis for the Binding Affinity. (The baseline is the execution using a single node)}
    \label{fig:pmx_weak_scaling}
\end{figure}

\begin{figure}[ht!]
    \centering
    \includegraphics[width=1\linewidth]{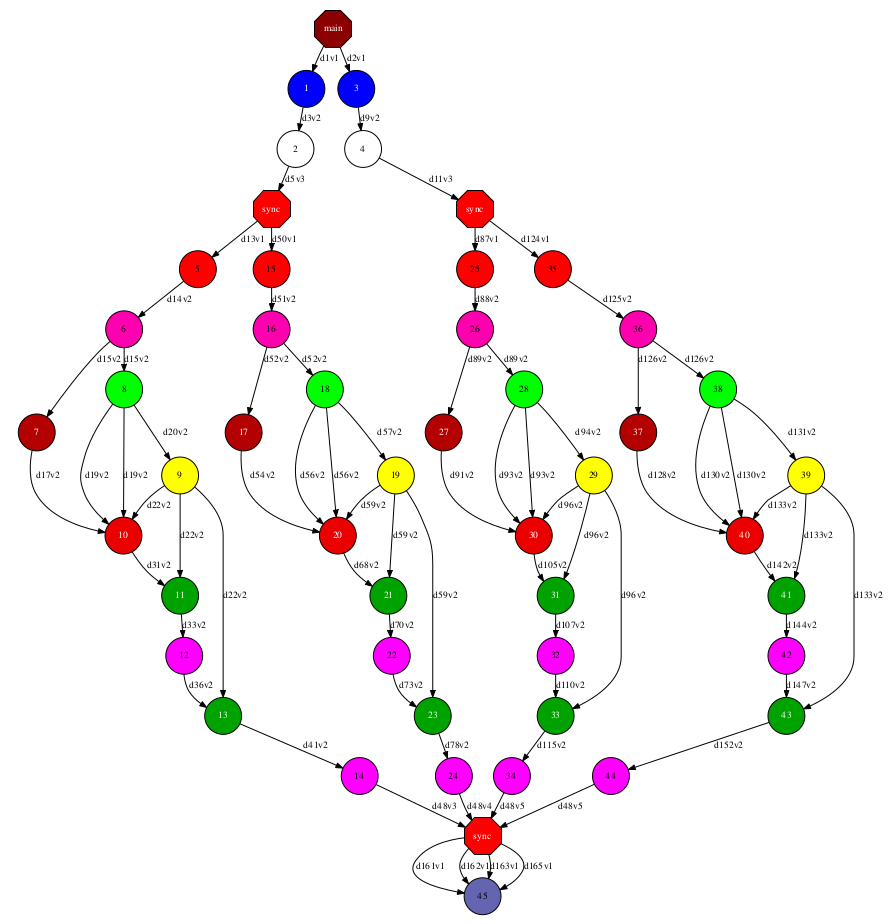}
    \caption{Generated task graph for 2
    structures per ensemble. Nodes in the graph represents tasks identified by the Workflow Management System and arrows represent the detected data dependencies}
    \label{fig:pmx_graph}
\end{figure}

To complete the scalability analysis, we have designed a weak scaling experiment. In this case, we have used the same Binding Affinity workflow and the same system (SARS-Cov-2 spike protein) but the number of structures to evaluate is increased by the same scale as the resources used for the execution, expecting to have a constant execution time. We have measured the execution time and we have computed the efficiency. The results of this experiment are presented in Figure~\ref{fig:pmx_weak_scaling} where we can see the efficiency is degraded with the amount of resources. For 512 nodes, the largest number of nodes, we have an 80\% of efficiency. The speed-up and efficiency are diverging from the ideal case for two reasons: first, the workflow management overheads, such as task scheduling and data transfers, are growing with the number of tasks and nodes; and second, the Binding-affinity workflow is not completely embarrassingly parallel. Figure~\ref{fig:pmx_graph} shows an the generated graph for a simple case for 2 structures per ensemble. In this graph, we can observe that there is an initial phase (blue and white tasks) with a very limited parallelism (2 parallel tasks) and a the last phase whose execution is sequential.   

\subsection{Reliability}
\begin{figure*}[ht!]
    \centering
    \includegraphics[width=0.45\linewidth]{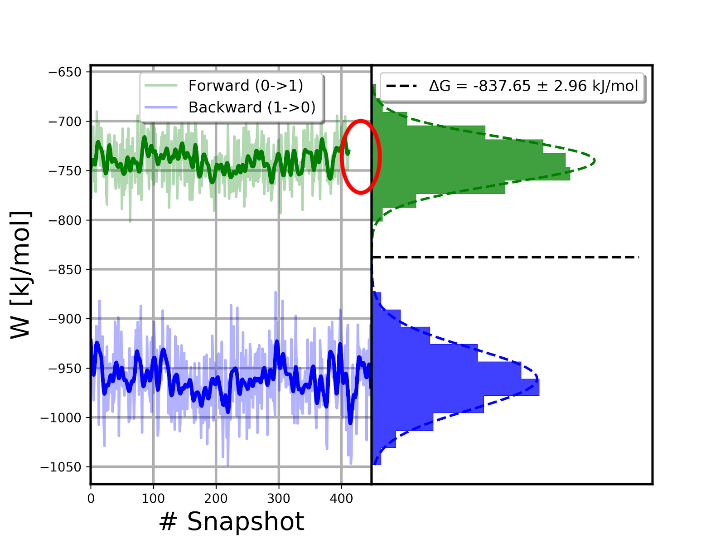}
    \includegraphics[width=0.45\linewidth]{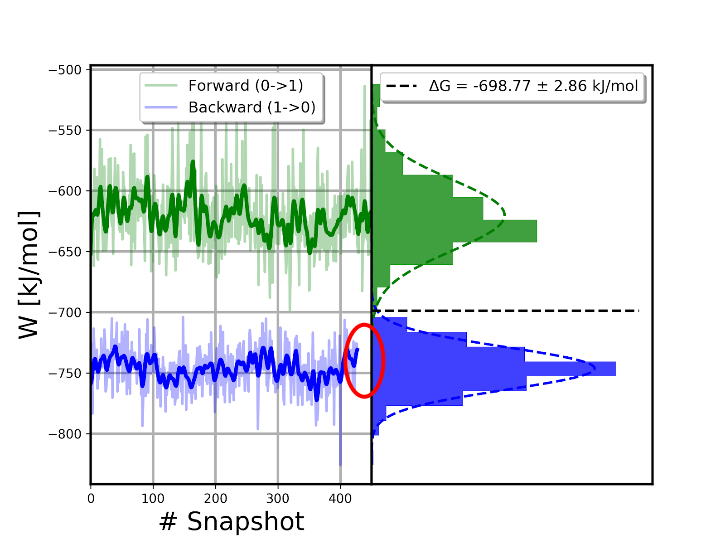}
    \caption{Binding Affinity workflow results with failures. The red circle indicates the simulations with empty results which are produced by the failures in the workflow}
    \label{fig:pmx_with_failures}
\end{figure*}

To validate the reliability features, we have executed the Binding Affinity workflow in the MareNostrum supercomputer. In this case, we have included the evaluation of some of the structures that produce failures in the simulations. Figure~\ref{fig:pmx_with_failures} shows the plot of two analyses performed by the workflow. The left-side of these plots contains a representation of the values obtained for the simulations performed with the different structures for the forward (green) and backward (blue) ensembles. Note that there are some empty values which correspond to the failed simulations (red circles). According to the policy indicated in the \textit{on\_failure} property, these failures have been ignored and empty values have been set, allowing the finalisation of the workflow execution without altering the results of the analysis.

\subsection{Malleability}
\begin{figure*}[tb!]
    \centering
    \includegraphics[width=1\linewidth]{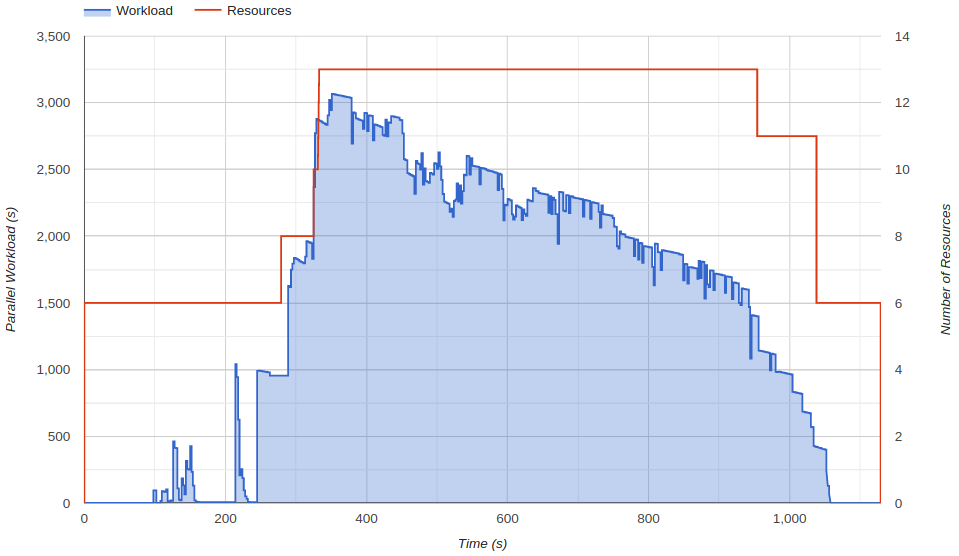}
    \caption{Malleability with the Binding Affinity workflow. Blue graph shows the parallel workload estimated by the runtime and the red line shows the number of resources assigned to the workflow execution.}
    \label{fig:md_elasticity}
\end{figure*}

To validate the malleability features of the proposed methodology, we have executed the Binding affinity workflow in the MareNostrum supercomputer with a malleable configuration, with 6 static computing nodes and 7 elastic computing nodes. Figure~\ref{fig:md_elasticity} shows the workload estimated by the runtime (blue) during the execution of the workflow, and the used amount of resources (red). During the first phase where setup tasks are quite small, the runtime decides to only use the static nodes. In the second phase where only the long simulations are pending the workload is increasing, so the runtime requests extra computing nodes that are released at the end of the execution, where the pending tasks can be finished with the static resources.

\section{Conclusions}
\label{sec:conclusions}
The paper has presented a methodology to reduce the gap between biomolecular research and the high performance computing world. The motivation for this work comes from the analysis, performed in the context of the BioExcel project, of the current situation around the execution of biomolecular workflows in supercomputing facilities, and it has developed around two pillars: usability and efficiency. 
The methodology consists of the combination of two main components: First, a python library of platform agnostic building blocks for molecular dynamics (BioBB) has been used to address the usability requirement. A wide variety of complex pipelines can be developed as simple python scripts. The second component of the methodology is the PyCOMPSs task based programming model. It allows to convert python scripts in parallel workflows whose execution is distributed in different computing nodes with a minimal impact on the code (just requiring to add simple annotation on top of the BioBB method definitions). 

We have described how the proposed methodology supports different types of simulations and adapts its execution to the available computing resources as well as other features like reliability and malleability. The validation of these features have been performed with two workflows implemented using the proposed methodology. These workflows have been executed in different premises with different processors to see how the execution is making use of the available hardware in each case. We have also verified that the workflow execution is finished even when some tasks are failing and the runtime system can adapt the infrastructure to the workload generated by the workflow execution. Finally, a scalability analysis has been performed executing the workflow with up to 65,536 cores, demonstrating that the workflows developed with this methodology can be easily scaled to a large number of nodes and that the optimal configuration of the execution parameters can be obtained without modifying the user code. 

Regarding future work, we will work on simplifying the development of more complex workflows and extending the BioBB library with the introduction of data analytics in the workflows. In particular, we will investigate how to couple the adoption of High Performance Data Analytics (HPDA) with High Performance Computing (HPC) techniques. The use of HPC parallel processing to run powerful data analysis software tools opens the possibility to examine massive datasets within a reasonable time. A set of HPDA building blocks will be developed, starting with a recently developed library for distributed computing integrated on top of the PyCOMPSs framework and focused on machine learning\cite{dislib}.


\section*{ACKNOWLEDGMENT}

This work has been supported by Spanish Ministry of Science and Innovation MCIN/AEI/10.13039/501100011033 under contract PID2019-107255GB-C21, by the Generalitat de Catalunya under contracts 2017-SGR-01414 and 2017-SGR-1110, by the European Commission through the BioExcel Center of Excellence (Horizon 2020 Framework program) under contracts 823830, and 675728. This work is also partially supported by the CECH project which has been co-funded with 50\% by the European Regional Development Fund under the framework of the ERFD Operative Programme for Catalunya 2014-2020, with a grant of 1.527.637,88€




\bibliographystyle{IEEEtran}
\bibliography{bibliography}


\end{document}